\documentclass[aps,prb,twocolumn,floatfix]{revtex4}
\usepackage{graphicx,bm}
\usepackage{comment}
\newcommand{\nn}{\nonumber\\}

\newcommand{\midb}[1]{\left[ #1 \right]}

\newcommand{\half}{{1\over 2}}
\begin{document}
\pagestyle{plain}
\title{Boltzmann Test of Slonczewski's Theory of Spin Transfer Torque}
\author{Jiang Xiao and A. Zangwill}
\affiliation{School of Physics, Georgia Institute of
Technology, Atlanta, GA 30332-0430}
\author{M. D. Stiles}
\affiliation{National Institute of Standards and
Technology, Gaithersburg, MD 20899-8412}
\begin{abstract}
    We use a matrix Boltzmann equation formalism to test the accuracy of Slonczewski's theory of spin-transfer torque in thin-film heterostructures where a non-magnetic spacer layer separates two non-collinear ferromagnetic layers connected to non-magnetic leads.  When applicable, the model predictions for the torque as a function of the angle between the two ferromagnets  agree extremely well with the torques computed from a Boltzmann equation calculation. We focus on asymmetric structures  (where the two ferromagnets and two leads are not identical) where the agreement pertains to a new analytic formula for the torque derived by us using Slonczewski's theory. In almost all cases, we can predict the correct value of the model parameters directly from the geometric and transport properties of the multilayer.  For some asymmetric geometries, we predict a new mode of stable precession that does not occur for the symmetric case studied by Slonczewski. 
\end{abstract}
\date{\today}
\maketitle
\begin{figure}
\centering
    \includegraphics[scale=0.6]{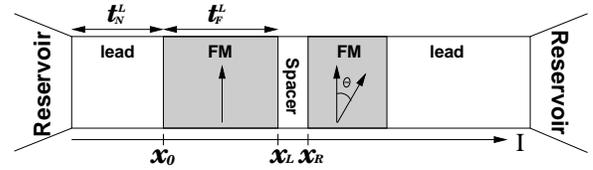}
    \caption{Schematic of a five-layer spin-valve. A non-magnetic spacer layer
    separates two ferromagnetic layers whose magnetizations are inclined from one another by an angle
    $\theta$. A non-magnetic lead connects each ferromagnet to an
    electron reservoir.}
    \vspace{-0.5cm}
    \label{fig:geometry}
\end{figure}

In 1996, Slonczewski\cite{Slonczewski:1996} and Berger\cite{Berger:1996} predicted that an electric current flowing through a magnetic multilayer can exert a spin-transfer torque on the magnetic moments of the heterostructure. This torque can produce stable magnetic precession and/or magnetic reversal, both of which have been widely studied experimentally\cite{Katine:2000}
and theoretically.\cite{Sun:2000}
Figure~\ref{fig:geometry} illustrates a  common geometry known as a ``spin-valve'' 
where a non-magnetic spacer layer separates a thick ``pinned'' ferromagnetic layer from a thin ``free'' ferromagnetic layer. Non-magnetic leads connect the ferromagnets to electron reservoirs.  

Slonczewski\cite{Slonczewski:2002} developed a theory of spin-transfer torque that combines a density matrix description of the spacer layer with a circuit theory description of the remainder of the structure. He worked out the algebra for the case where the spacer layer is thin and the spin valve is symmetric (identical ferromagnets and leads) and found the torque $L_S$ to be the same on the left and right spacer/ferromagnet interfaces. As a function of the angle between the two ferromagnets,   
\begin{equation}
\label{torque}
    L_S(\theta)
    ={\hbar I  \over 2e}{P\Lambda^2\sin\theta \over (\Lambda^2+1) +(\Lambda^2-1)\cos\theta}.
\end{equation}
In this formula,  $I$ is the total current that flows through the structure, 
\begin{equation}
\label{P}
    P
    ={\textstyle{1\over2}(R_\downarrow-R_\uparrow)\over \textstyle{1\over2}(R_\downarrow+R_\uparrow)}
    ={r\over R} 
    \hspace{20pt}{\rm and}\hspace{20pt} 
    \Lambda^2=GR.
\end{equation}
$R_\uparrow$ and $R_\downarrow$ are {\it effective} resistances experienced by spin up and spin down electrons between the reservoir and the spacer layer. The conductance $G=Ae^2k_F^2/4\pi^2 \hbar$, where $A$ is the cross-sectional area of the device.

In this paper, we solve Slonczewski's equations for the general asymmetric case and derive formulae for the torques $L_L(\theta)$ and $L_R(\theta)$ on the left ($x=x_L$) and right ($x=x_R$) spacer/ferromagnet interfaces in Figure~\ref{fig:geometry}. We then compare these formulae with numerical results for the torque obtained from a matrix Boltzmann equation.\cite{SZ:2002b} From this comparison, we are able to identify the physical origin and systematic behavior of the effective resistances $R_\uparrow$ and $R_\downarrow$ in the interplay of inelastic scattering, spin-flip scattering, and interface scattering.  For some asymmetric geometries, a previously unsuspected feature of the torque leads us to predict a new mode of stable precession that does not occur for the symmetric case. 

The electric current in a spin valve can be written as the sum of spatially varying currents carried by up and down\cite{updown} spin electrons: $I=I_\uparrow(x)+I_\downarrow(x)$. The corresponding spin-current\cite{caveat} is $Q(x)=I_\downarrow(x) - I_\uparrow(x)$. The up and down spin voltage (chemical potential) drops along the spin valve are $V_\uparrow(x)$ and $V_\downarrow(x)$. Slonczewski\cite{Slonczewski:2002} contracted the function  $I_\uparrow(x)$ to the values $I^L_\uparrow=I_\uparrow(x_L)$ and $I^R_\uparrow=I_\uparrow(x_R)$ and similarly for $I_\downarrow(x)$. He also writes $\Delta V_R=V_\uparrow(x_R)-V_\downarrow(x_R)$ for the up-down difference in the voltage drop from the right reservoir to a point in the spacer infinitesimally close to the interface between the spacer and the right ferromagnet. $\Delta V_L$ is defined similarly. These quantities are directly proportional to the spin accumulation at $x_R$ and $x_L$, respectively.

With this model, Slonczewski wrote down (but did not completely solve) all the equations needed for the asymmetric geometry. In our notation, two of his linear equations relate the voltage drop differences to the spin currents: 
\begin{eqnarray}
\label{eqn:JW}
    0 &=& \Delta V_L (1+\cos^2\theta) - G^{-1} Q_L \sin^2\theta-2\Delta V_R \cos \theta \nn
    0 &=& Q_L(1+\cos^2\theta) - G \Delta V_L \sin^2\theta - 2Q_R \cos \theta . 
\end{eqnarray}
Two additional equations parameterize the voltage drop differences in terms of effective resistances $R_L$, $R_R$, $r_L$ and $r_R$: \vspace{-1em}
\begin{eqnarray}
\label{eqn:JWW}
    \Delta V_L &=& Q_L R_L + I r_L \nn
    \Delta V_R &=&  - Q_R R_R - I r_R. 
\end{eqnarray}
Finally, Slonczewski derived expressions for the interfacial torques. At $x=x_R$, the torque is
\begin{equation}    
\label{righttorque}
    L_R = {\hbar\over 2e}{Q_R\cos\theta - Q_L\over\sin\theta}.
\end{equation}
The torque $L_L$ at the $x=x_L$ interface is (\ref{righttorque}) with $Q_R$ and $Q_L$ exchanged.

It is straightforward to solve the four equations (\ref{eqn:JW}) and (\ref{eqn:JWW}) for the four  unknowns, $Q_L$, $Q_R$, $\Delta V_L$, and $\Delta V_R$. From these, we find $L_R$ from (\ref{righttorque}) to be
\begin{equation}
\label{eqn:torque2}
    L_R = {\hbar \over 2e}{I \sin\theta\over A}
        \left[ {q_+ \over A + B \cos\theta} + {q_- \over A - B \cos\theta} \right]
\end{equation}        
where        
\begin{eqnarray}        
\label{qAB}
    q_\pm &=& \half \left[ P_L\, \Lambda_L^2 \sqrt{\Lambda_R^2 + 1 \over \Lambda_L^2 + 1}
                          \pm P_R \,\Lambda_R^2 \sqrt{\Lambda_L^2 - 1 \over \Lambda_R^2 - 1}\right]\nn
    A   &=& \sqrt{(\Lambda_L^2 + 1)(\Lambda_R^2 + 1)}  \nn
    B   &=& \sqrt{(\Lambda_L^2 - 1)(\Lambda_R^2 - 1)}.
\end{eqnarray}
The parameters $P_L$, $P_R$, $\Lambda_L$ and $\Lambda_R$ are defined in terms of $R_L$, $R_R$, $r_L$ and $r_R$ as $P$ and $\Lambda$ are defined in (\ref{P}) in terms of $R$ and $r$. For the symmetric case, $\Lambda_L = \Lambda_R=\Lambda$ and $P_L = P_R=P$. This makes $q_- = 0$ and (\ref{eqn:torque2}) reduces to  Slonczewski's formula (\ref{torque}) with $L_R=L_L=L_S$. We will see later that the term proportional to $q_-$ in (\ref{eqn:torque2}) can affect the magnetization dynamics of the spin valve in a qualitative way.

To test (\ref{eqn:torque2}), we computed the torque on each spacer/ferromaget interface using a Boltzmann equation formalism.\cite{SZ:2002b} The two ferromagnetic moments in Figure~\ref{fig:geometry} are not collinear, so there is no natural spin quantization axis in the spacer layer. Therefore, we expand the  deviation of the semi-classical electron occupation function from its equilibrium value in a basis of Pauli spin matrices:
\begin{eqnarray}
{\bf g}({\bf k},{\bf r})&=&g_0({\bf k},{\bf r})\left(\begin{array}{cc}
  1 & 0 \\
  0 & 1 \\
\end{array} \right)+g_x({\bf k},{\bf r})\left(\begin{array}{cc}
  0 & 1 \\
  1 & 0 \\ 
\end{array} \right) \nn &+&g_y({\bf k},{\bf r})\left(\begin{array}{cc}
  0 & -i \\
  i & 0 \\
\end{array} \right)  
 + g_z({\bf k},{\bf r})\left(\begin{array}{cc}
  1 & 0 \\
  0 & -1 \\
\end{array} \right)
\end{eqnarray}
In the spacer layer, each of $g_0$, $g_x$, $g_y$, and $g_z$ satisfies a linear Boltzmann equation that takes account of the driving current and of resistive and spin-flip scattering in each material layer. In each ferromagnet and in the adjacent lead, it is sufficient to use $g_0$ and $g_z$ referenced to the fixed direction
of magnetization in each.  We assume a spherical Fermi surface for each material. We adopt a one-dimensional approximation ${\bf g}({\bf k},{\bf r})={\bf g}({\bf k},x)$ and stitch together the solutions in each layer with matching conditions that reflect the complex, spin- and wave vector-dependent  reflection and transmission amplitudes of each interface. The latter are determined from a previously published parameterization.\cite{Stiles:2000} 

Electrons that enter a non-magnetic lead from the adjacent reservoir are assumed to have an equilibrium  occupation function.
For our numerical work, we have chosen material and geometry parameters typical of experiments performed on Cu/Co/Cu/Co/Cu spin valves. The specific numerical values used can be found in Ref.~\onlinecite{SZ:2002b}. With the final occupation function in hand, it is straightforward to compute the spatially varying voltage, spin  accumulation, spin current, and spin-transfer torque. Details will be published elsewhere.\cite{Xiao}
\begin{figure}
\centering
    \includegraphics[scale=0.7]{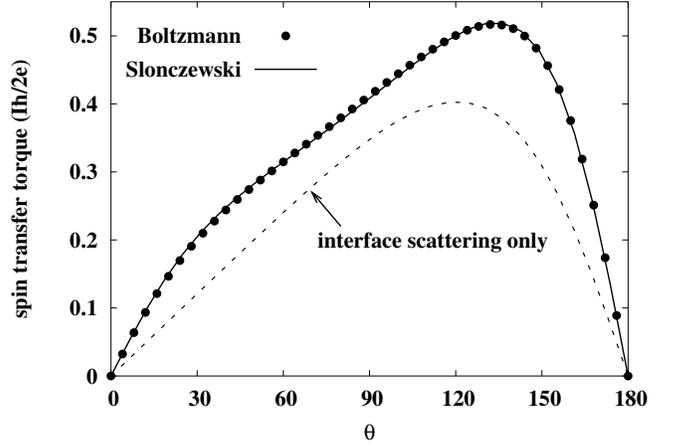}
    \caption{Spin-transfer torque at $x=x_R$ for a spin-valve with layer thicknesses 5\,nm/40\,nm/1\,nm/1\,nm/180\,nm. Solid circles are Boltzmann equation results. Solid curve is Eq.~\ref{eqn:torque2} derived from Slonczewski's theory. Dashed curve is Eq.~\ref{eqn:torque2} with all bulk scattering removed. }
    \vspace{-0.5cm}
    \label{fig:boltzmann-extslon}
\end{figure}

The filled circles in Figure~\ref{fig:boltzmann-extslon} show a typical Boltzmann result for the spin-transfer torque at $x=x_R$ as a function of the angle $\theta$ between the two ferromagnets for an asymmetric geometry. The solid curve in   Figure~\ref{fig:boltzmann-extslon} is  the same quantity, $L_R(\theta)$, computed from (\ref{eqn:torque2}). The agreement is excellent, as it is for essentially all other geometries we have studied with thin spacer layers. For comparison, the dashed curve is the Slonczewski torque  (\ref{eqn:torque2}) with the bulk scattering  removed. This ``interface-only'' situation is manifestly symmetric ($q_-=0$).

For the solid curve plotted in  Figure~\ref{fig:boltzmann-extslon}, the relative importance of the two terms in (\ref{eqn:torque2}) is  $q_-/q_+ \simeq 0.27$. We find that this ratio does not exceed about $0.5$ for physically sensible geometries. We will address the qualitative consequences of $q_-\neq 0$ at the end. First, we describe our method to determine the torque parameters $\Lambda_R$, $\Lambda_L$, $P_R$, and $P_L$ that produced the solid curve in Figure~\ref{fig:boltzmann-extslon}. We begin by writing an exact expression for the voltage drop difference $\Delta V_L$ in (\ref{eqn:JWW}):
\begin{eqnarray} 
\label{inter}
    \Delta V_L
    & = &\int\limits_{-\infty}^{x_L} dx [I_\downarrow(x)\rho_\downarrow(x)-I_\uparrow(x)\rho_\uparrow(x)] \nn
    & = &\int\limits_{-\infty}^{x_L} dx\, [\hat{I}_\downarrow(x)-\hat{I}_\uparrow(x)]\,\bar{\rho}(x)
\end{eqnarray}
$\hat{I}_\sigma(x)$ is $I_\sigma(x)$ {\it minus} the corresponding current in an infinite bulk sample of the material present at point $x$. The  average resistivity in (\ref{inter}) is $\bar{\rho}=(\rho_\uparrow + \rho_\downarrow)/2$. We will also need the resistivity difference $\Delta \rho = (\rho_\downarrow - \rho_\uparrow)/2$. Both $\bar{\rho}$ and $\Delta \rho$ contain delta functions at the four non-magnet/ferromagnet interfaces to take account of spin-dependent interface scattering.

To make progress, we use a drift-diffusion result\cite{Stiles:2004} to the effect that $\hat{I}_\uparrow(x)$ and
$\hat{I}_\downarrow(x)$ decay exponentially close to the $x=x_0$ interface (Figure~\ref{fig:geometry}) in both directions. The decay length is the spin-flip length in each material, whether ferromagnet (F) or non-magnet (N). 
In that case, an approximate expression for (\ref{inter}) is
\begin{eqnarray}
\label{eqn:deltaw}
    \Delta V_L 
    &=& Q_0\bar{\rho}_{\rm N}\, d_{\rm N}^L + Q_0\bar{\rho}_{\rm F}\, d_{\rm F}^L 
     +  I \Delta \rho_{\rm F}\, d_{\rm F}^L \nn &+& \Delta V_I +\Delta V_C.
\end{eqnarray}
In (\ref{eqn:deltaw}), $Q_0=Q(x_0)$, $\Delta V_I$ and $\Delta V_C$ are voltage drops at the internal
interfaces and at the reservoir contact, and 
\begin{eqnarray}
\label{length}
    d^L_{\rm F} &= &l_{\rm sf}^{\rm F}\midb{1 - \exp(-t_{\rm F}^L/l_{\rm sf}^{\rm F})} \nn
    d^L_{\rm N} &= &l_{\rm sf}^{\rm N}\midb{1 - \exp(-t_{\rm N}^L/l_{\rm sf}^{\rm N})}. 
\end{eqnarray}
The effective lengths (\ref{length}) appear  because, due to spin-flip scattering, only electrons within $d_{\rm F}$ or $d_{\rm N}$ of the ferromagnetic interfaces can accommodate the dissimilar spin-currents characteristic of the ferromagnets and the non-magnets in equilibrium.  

The relationship between $Q_0$ and $Q_L$ is non-trivial\cite{Stiles:2004} except when the ferromagnet is very thin ($t_{\rm F}^{\rm L} \ll l_{\rm sf}^{\rm F}$). In that case, $Q_0 \simeq Q_L$, and we can connect (\ref{eqn:deltaw}) to (\ref{eqn:JWW}) and (\ref{eqn:torque2}) to get
\begin{eqnarray}
\label{eqn:torque}
    \Lambda^2_L &=& 
    G \,(\bar{\rho}_{\rm N} \,d_{\rm N}^L + \bar{\rho}_{\rm F}\, t_{\rm F}^L + \bar{R}_I + \bar{R}_C ) \nn
    P_L &=&  
    \Lambda_L^{-2}\,(\Delta\rho_{\rm F}\, t_{\rm F}^L + \Delta R_I). 
\end{eqnarray}
These two formulae (and similar ones for $\Lambda_R$ and $P_R$), together with  (\ref{eqn:torque2}) and
(\ref{qAB}) are the principal results of this paper. 
We used (\ref{eqn:torque}) to compute the solid curve in Figure~\ref{fig:boltzmann-extslon}. The interface resistance $GR_I\approx 0.97$ and contact resistance $GR_C \approx 1.1$ were extracted from the Boltzmann solution.  $GR_I$ differs from the  experimental value by about $15\%$.\cite{Bass}  

\begin{figure}
\centering
    \includegraphics[scale=0.33]{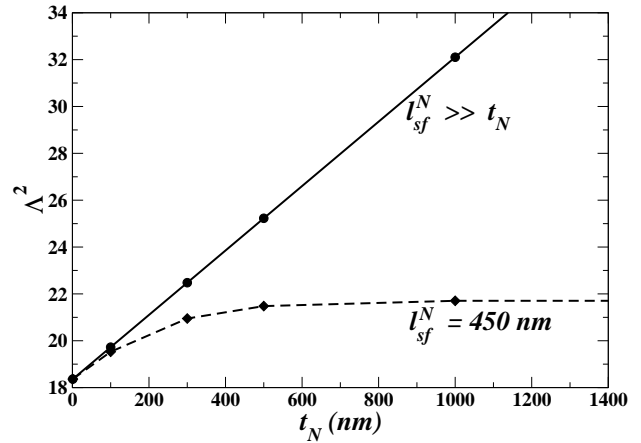}
    \caption{Torque parameter $\Lambda^2$ as a function of $t_{\rm N}$ for large and small values of
    $l_{\rm sf}^{\rm N}$.}
    \label{fig:spinflip}
\end{figure}
The parametrization (\ref{eqn:torque2}) is well suited to study the behavior of $L(\theta)$ when we vary the geometry of the magnetic heterostructure and the material parameters in our Boltzmann calculations. For convenience, we did this for symmetric geometries. Figure~\ref{fig:spinflip} confirms that $\Lambda^2$ is a linear function of $t_{\rm N}$ when $l_{\rm sf}^{\rm N} \gg t_{\rm N}$ but saturates when $t_{\rm N} \sim l_{\rm sf}^{\rm N}$.  Interestingly, the saturated value of $\Lambda^2$ varies linearly with  
$l_{\rm sf}^{\rm N}-l^{\rm N}$ ($l^{\rm N}$ is the inelastic scattering length) rather than with $l_{\rm sf}^{\rm N}$ as predicted by (\ref{length}). For long leads, this can be understood from the fact that conventional resistive scattering is needed to build up non-equilibrium spin accumulation in the non-magnet while spin-flip scattering works to return the non-magnet to equilibrium. 
\begin{figure}
\centering
    \includegraphics[scale=0.33]{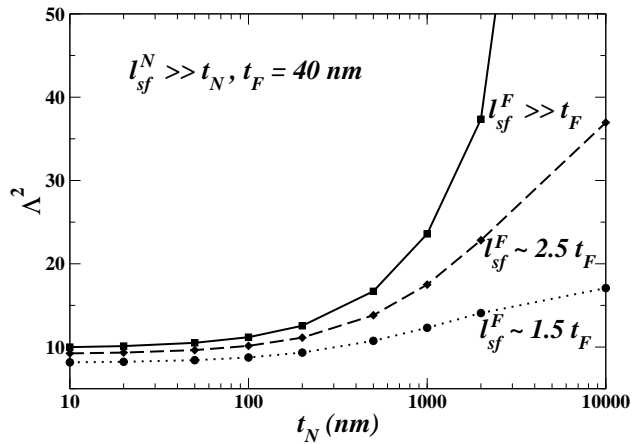}
    \caption{Torque parameter $\Lambda^2$ as a function of $t_{\rm N}$ for different values of $l_{\rm sf}^{\rm F}$. In all cases, $l_{\rm sf}^{\rm N}=\infty$. The curve for $l_{\rm sf}^{\rm F}\gg t_{\rm F}$ corresponds to $\Lambda^2 \propto t_{\rm N}$.}
    \vspace{-0.5cm}
    \label{fig:Fflip} 
\end{figure}
\begin{figure*}
\centering
    \includegraphics[scale=0.46]{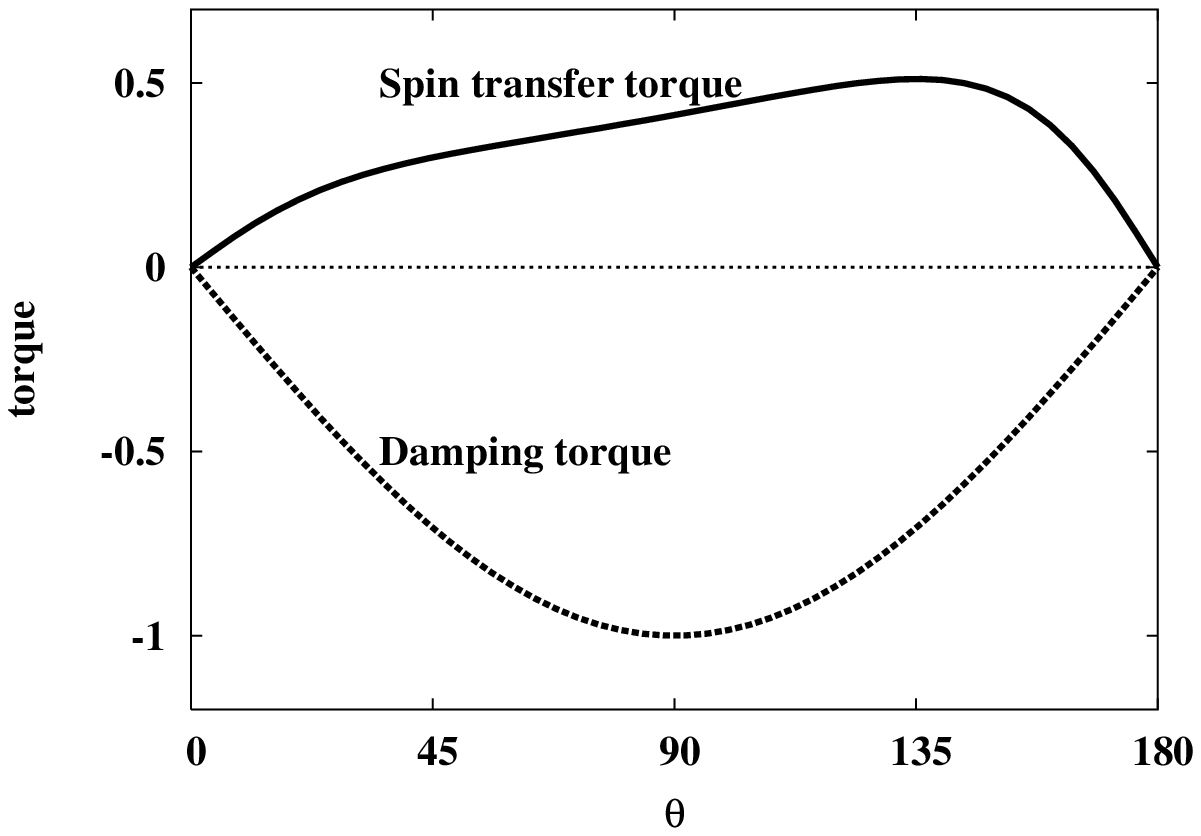}
    \includegraphics[scale=0.46]{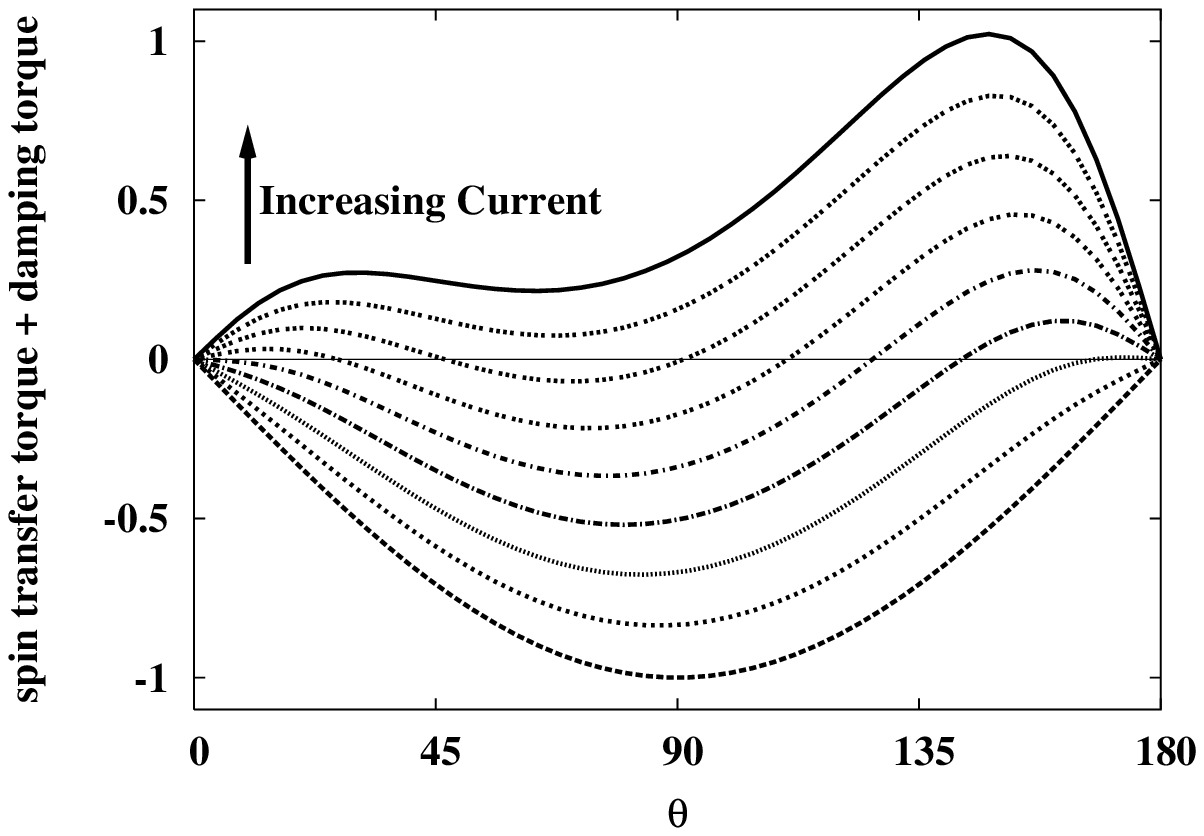}
    \includegraphics[scale=0.46]{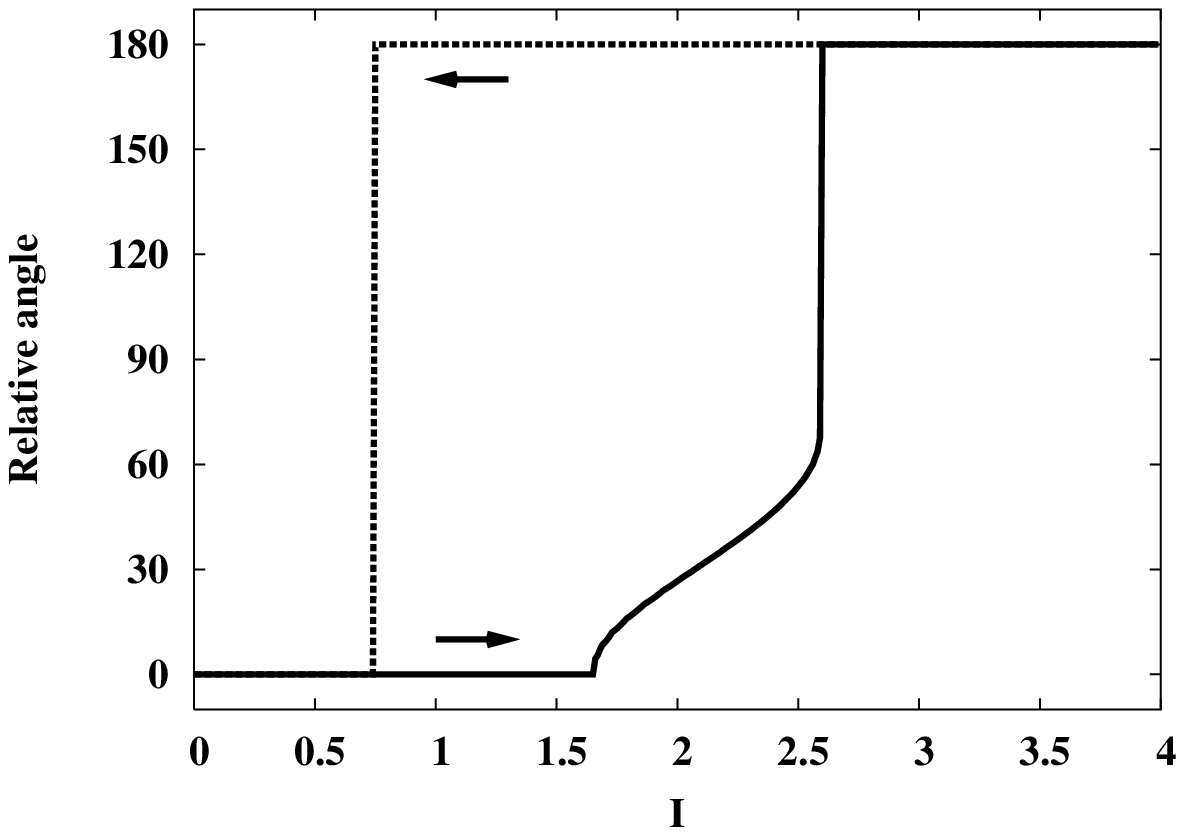}
    \caption{Left: Spin transfer torque and damping torque for a 1\,nm/40\,nm/1\,nm/1\,nm/1000\,nm spin valve ($q_-/q_+\approx 0.36$);  Middle: the total torque as the current increases; Right: the angle between two ferromagnetic moments as a function of current $I$ in arbitrary units.}
    \vspace{-0.5cm}
    \label{fig:torquecompare}
\end{figure*}

Figure~\ref{fig:Fflip} shows the variation of $\Lambda^2$ with lead length for different values of the spin-flip length in the ferromagnet. This calculation puts $l_{\rm sf}^{\rm N}\to \infty$, so we expect from (\ref{length}) that $\Lambda^2 \propto t_{\rm N}$. This is indeed
the case when $l_{\rm sf}^{\rm F} \gg t^{\rm F}$. However, when the $l_{\rm sf}^{\rm F}$ is comparable (or less than) the ferromagnetic layer thickness, the torque parameter saturates.
This is a signal that our approximation $Q_0 \simeq Q_L$ has broken down. In this limit, fast spin-flipping in the ferromagnet reduces $Q_0$ to a value much less than $Q_L$. When that is the case, relatively little spin-flip volume in the non-magnet is needed to reduce the spin current in the non-magnet to zero (its equilibrium value).  
A characteristic difference between $L_S(\theta)$ in (\ref{torque}) for a symmetric geometry and $L_R(\theta)$ in  (\ref{eqn:torque2}) for an asymmetric geometry can be seen from the difference between the dashed curve and the solid curve in Figure~\ref{fig:boltzmann-extslon}. The former has a bump (maximum) in the interval $\pi/2 < \theta < \pi$ only. The latter has a small additional bump in the interval $0 < \theta < \pi/2$ that comes from the $q_-$ term.\cite{cvS} This small change is enough to produce stable magnetization precession for some asymmetric geometries. 

Consider a spin valve in the presence of an external magnetic field aligned with the magnetization of the thick ferromagnet. If we ignore shape anisotropy and lattice anisotropy, the total torque acting on the thin ferromagnetic film when electrons flow from right to left in Figure~\ref{fig:geometry} is the sum of the spin-transfer torque $L_R(\theta)$ and a Gilbert damping torque $\gamma H \sin{\theta}$ (left panel of Figure~\ref{fig:torquecompare}). As the current increases, the spin-transfer torque increases and eventually destablizes an initial state with parallel moments. Stable precession occurs at angles where the total torque changes from positive to negative (middle panel of Figure~\ref{fig:torquecompare}).
When the total torque becomes everywhere positive, the system abruptly switches to the anti-parallel configuration (right panel of Figure~\ref{fig:torquecompare}). There is no regime of stable precession if the zero-current state is anti-parallel. 

In summary, we have shown that Slonczewski's theory of spin-transfer torque in spin-valves can reproduce the results of Boltzmann equation calculations when the non-magnetic spacer layer is thin. When the ferromagnetic layers are also thin, the parameters of the theory can be calculated from first principles.  
One of us (J.X.) acknowledges support from the National Science
Foundation under grant DMR-9820230.\\


\begin{thebibliography}{99}
\bibitem{caveat}The spin current $Q_{\alpha\beta}$ is a second rank tensor.  Here, we need only one spatial index (the direction of current flow) and
one spin index (the local direction of spin polarization), so a scalar $Q$ is sufficient.
\end{thebibliography}
\end{document}